\pdfoutput=1 
\documentclass[journal=nalefd,manuscript=letter]{achemso}

\usepackage[version=3]{mhchem} 
\usepackage{physics}
\usepackage{amssymb}

\author{Tobias M. Krieger}
\email{tobias.krieger@jku.at}
\author{Christian Weidinger}
\author{Thomas Oberleitner}
\author{Gabriel Undeutsch}
\affiliation{Institute of Semiconductor and Solid State Physics, Johannes Kepler University Linz, Altenberger Straße 69, 4040 Linz, Austria}
\author{Michele B. Rota}
\affiliation{Dipartimento di Fisica, Sapienza University of Rome, Piazzale Aldo Moro 5, 00185 Rome, Italy}
\author{Naser Tajik}
\author{Maximilian Aigner}
\affiliation{Institute of Semiconductor and Solid State Physics, Johannes Kepler University Linz, Altenberger Straße 69, 4040 Linz, Austria}
\author{Quirin Buchinger}
\affiliation{Julius-Maximilians-Universität Würzburg, Physikalisches Institut, Lehrstuhl für Technische Physik, Am Hubland, 97074 Würzburg, Deutschland}
\author{Christian Schimpf}
\author{Ailton J. Garcia Jr.}
\author{Saimon F. Covre da Silva}
\affiliation{Institute of Semiconductor and Solid State Physics, Johannes Kepler University Linz, Altenberger Straße 69, 4040 Linz, Austria}
\author{Sven Höfling}
\author{Tobias Huber-Loyola}
\affiliation{Julius-Maximilians-Universität Würzburg, Physikalisches Institut, Lehrstuhl für Technische Physik, Am Hubland, 97074 Würzburg, Deutschland}
\author{Rinaldo Trotta}
\affiliation{Dipartimento di Fisica, Sapienza University of Rome, Piazzale Aldo Moro 5, 00185 Rome, Italy}
\author{Armando Rastelli}
\affiliation{Institute of Semiconductor and Solid State Physics, Johannes Kepler University Linz, Altenberger Straße 69, 4040 Linz, Austria}
\email{armando.rastelli@jku.at}

\title
  {Post-fabrication tuning of circular Bragg resonators for enhanced emitter-cavity coupling}

\begin{document}

\begin{abstract}
Solid-state quantum emitters embedded in circular Bragg resonators are attractive due to their ability to emit quantum states of light with high brightness and low multi-photon probability. As for any emitter-microcavity system, fabrication imperfections limit the spatial and spectral overlap of the emitter with the cavity mode, thus limiting their coupling strength.
Here, we show that an initial spectral mismatch can be corrected after device fabrication by repeated wet chemical etching steps. We demonstrate $\sim$16~nm wavelength tuning for optical modes in AlGaAs resonators on oxide, leading to a 4-fold Purcell enhancement of the emission of single embedded GaAs quantum dots. Numerical calculations reproduce the observations and suggest that the achievable performance of the resonator is only marginally affected in the explored tuning range. 
We expect the method to be applicable also to circular Bragg resonators based on other material platforms, thus increasing the device yield of cavity-enhanced solid-state quantum emitters.
\end{abstract}

Bright sources of non-classical light play a crucial role in the development of quantum technologies such as quantum communication, and quantum information processing~\cite{Kimble2008,Gisin2007,Fedrizzi2009}. 
Over the past decades, various schemes for generating single photons~\cite{Meyer-Scott2020} and entangled photon pairs~\cite{Orieux2017} have emerged. 
Among these schemes, sources based on spontaneous parametric down conversion (SPDC) are commonly used for heralded single photons and entangled photon pairs with a near-unity degree of entanglement~\cite{Kwiat1999}. 
However, the stochastic nature of photon generation of such sources imposes fundamental limitations on their maximum brightness~\cite{Jons2017}.

Solid-state emitters are promising alternatives for single photon and entangled photon sources~\cite{Huber2018}, as they combine optical quality similar to atomic emission~\cite{Kuhn2002,Schweickert2018, Crocker2019} with compact nanoscale integration, enabling the use of established manufacturing processes of the host system~\cite{Aharonovich2016}. 
Nevertheless, solid-state systems present challenges, including an inhomogeneous distribution of emission properties among multiple emitters, as well as homogeneous and inhomogeneous broadening of the emission linewidth, which reduces photon indistinguishability. 
Moreover, for host materials with a large refractive index, the extraction efficiency of photons from the solid-state is limited due to total internal reflection.

To address these challenges, circular Bragg grating resonators (CBRs), also known as ``bullseye cavities" or ``bullseye antennas"~\cite{Davanco2011} emerged as very appealing structures enabling high extraction efficiency over a large frequency range and Purcell enhancement of quantum emitters coupled to optical resonator modes. 
The bullseye design has been successfully implemented in various emitter systems, including integration with III-V semiconductor quantum dots (QDs)~\cite{Ates2012, Liu2019, Wang2019, Kolatschek2021,Xia2021, Xu2022, Barbiero2022, Rota2022}, emitters in GaN layers~\cite{Meunier2023}, color centers in hexagonal boron nitride~\cite{Froch2021}, nitrogen vacancy centers in diamond~\cite{Li2015}, as well as plasmonic coupling of bullseye antennas to N and Si vacancy centers in nanodiamonds~\cite{Andersen2018,Waltrich2021}, emitters in WSe$_2$ monolayers~\cite{Iff2021}, and colloidal QDs~\cite{Werschler2018,Abudayyeh2021}.

Although the low quality factor of the CBR allows some tolerance to spectrally match the cavity mode (CM) with integrated emitters, the realization of such devices remains challenging. 
Fabrication imperfections limit the accuracy of the spatial and spectral overlap between emitter and desired cavity mode, especially for short-wavelength emitters embedded in high refractive-index materials, where deviations of a few tens of nm represent already a significant fraction compared to the effective wavelength. 
This results in a low yield of deterministically patterned devices, while still requiring time-consuming pre-characterization of the emitters.
Several tuning approaches can be employed to match the emission energy of an emitter with a CM, including engineering of the strain-field~\cite{Zander2009, Sun2013, Hepp2020,Moczaa-Dusanowska2019,Rota2022} or the application of an electric field~\cite{Singh2022}.
Alternatively, tuning of the CM can be achieved through post-fabrication strategies, such as atomic layer deposition~\cite{Kiravittaya2011}, laser-assisted oxidation~\cite{Lee2009}, temperature~\cite{Benyoucef2008}, free-carrier absorption~\cite{Fushman2007}, or gas condensation at low temperatures~\cite{Mosor2005}. 
Wet chemical etching~\cite{Hennessy2005,Sunner2007} allows for broad-band tuning, but concerns may arise about possible degradation of the optical performance of the emitters and resonators, and about the structure integrity due to etching of the dielectric layer placed usually between CBR and backside reflector~\cite{Liu2019,Wang2019}.

In this study, we demonstrate that repeated wet chemical etching of the native oxides of GaAs and AlGaAs enables post-fabrication tuning of the CM wavelength by 31(3)~meV (16(2)~nm), resulting in spectral matching with the emission of an embedded GaAs QD obtained by droplet-etching epitaxy~\cite{DaSilva2021}. 
This tuning approach enables 4-fold Purcell enhancement -- here limited by inaccurate spatial overlap between emitter and cavity -- without compromising either the single photon purity, optical quality, or the structural integrity of the AlGaAs resonator on Al$_2$O$_3$ dielectric layer.

\begin{figure}
  \includegraphics{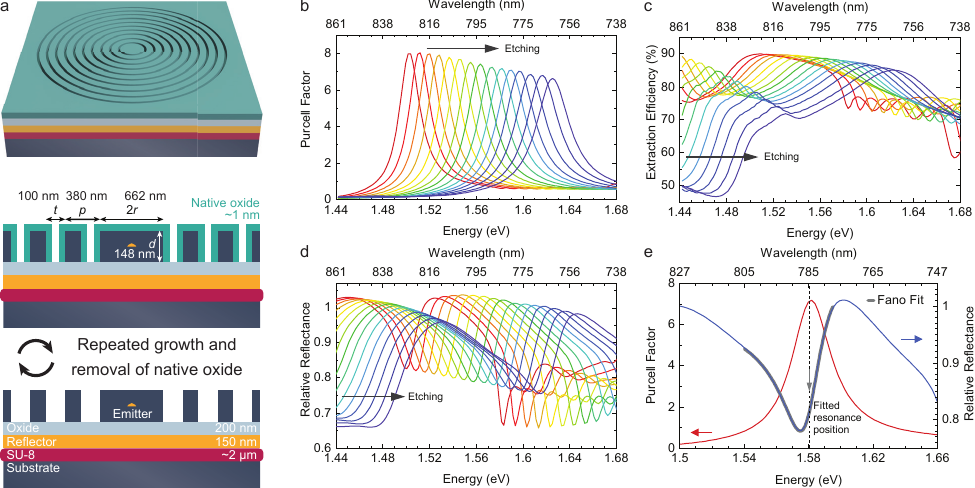}
  \caption{(a) Sketch representing the CBR structure with geometric parameters before processing and the repeated removal of the native oxide by wet chemical etching. FDTD simulation results of (b) the Purcell enhancement, (c) the extraction efficiency, and (d) the CBR reflectance relative to that of the surrounding planar areas for repeated etch steps as a function of photon energy and wavelength. e) Reflectance curve fitted with a Fano lineshape with marked resonance position $E_{\text{c}}$ compared to the corresponding spectrum of the Purcell factor.}
  \label{fig:1}
\end{figure}

The sample under investigation, as sketched in Figure~\ref{fig:1}(a), consists of a 140~nm thick Al\textsubscript{0.33}Ga\textsubscript{0.67}As membrane, containing a GaAs QD layer in the center, with 4~nm thick GaAs capping layers on top and bottom, and a back-reflector underneath. 
CBRs with three designs (d1, d2, d3), each characterized by a different radius $r$ of the central disc, are dry-etched deterministically on QD sites, exposing the side walls of Al\textsubscript{0.33}Ga\textsubscript{0.67}As to the ambient. 
Details of sample fabrication are provided in the [Supporting Information].
CBR structures on this sample exhibit limited spatial and spectral overlap with the QDs.
The QD emission displays a degree of polarization higher than 60~\%, which we attribute to a misplacement of the QDs from the center of the cavity~\cite{Peniakov2023}.
Furthermore, no Purcell enhancement was observable after fabrication due to a systematical blue shift, of about 26-49 meV (13-25nm), compared to the cavity mode (see below).

We now illustrate the concept of mode tuning by using finite-difference time-domain (FDTD) simulations for the CBR sketched in Figure~\ref{fig:1}(a) with an emitting dipole located at its center. (For details cf. to the section ``Simulation" in the [Supporting Information]). 
Figure~\ref{fig:1}(b-d) illustrate that by material removal from the CBR, i.e., by reducing the center disc radius $r$, decreasing the membrane thickness $d$, and increasing the trench width $t$, in steps of 1.5~nm (in total 21~nm), we expect a blue-shift in the spectral position of the maximum Purcell factor (Figure~\ref{fig:1}(b)). 
Simultaneously, the extraction efficiency peak (Figure~\ref{fig:1}(c)) is also blue shifted. 
The peak value of the extraction efficiency slowly decreases for increasing etch steps, as the structure continuously departs from the optimized design for a given wavelength (see also simulations of the quality factor in the [Supporting Information]). 
Nevertheless, the efficiency stays $\gtrsim$85~\% in a $\sim$20-40~meV~(10-20~nm) wide range around the CM position, so we can concentrate on shifting the CM position to achieve Purcell enhancement without worrying about significant intensity drops. 

In the experiment it is convenient to use reflectance spectroscopy to obtain the spectral position of the CM. 
To unambiguously find the spectral position of the Purcell enhancement-maximum, we simulate the reflectance spectrum, as shown in  Figure~\ref{fig:1}(d).
The asymmetry of the resulting reflectance dip arises from the Fano interference effect, which occurs in presence of two possible pathways of scattering events from a discrete state and from a continuum~\cite{Fano1961,Limonov2017}. 
Light that is directly scattered from the surface interferes with light that scatters resonantly coupled to the CM~\cite{Galli2009}. 
This effect is also observable in CBRs~\cite{Buchinger2023}.
The Fano lineshape $I(E)$ is given by 
\begin{equation}
  I(E) = B + A\frac{(q+\Omega)^2}{1 + \Omega^2}, \label{eqn:fano}
\end{equation}
with energy $E$, constants $A, B$ and $\Omega = 2(E-E_{\text{c}})/\Gamma_{\text{c}} $, where $E_{\text{c}}$ is the CM position and $\Gamma_{\text{c}}$ the resonance linewidth. 
The Fano parameter $q$ is the ratio of direct and resonant transition amplitudes of scattering events~\cite{Fano1961} and influences the asymmetry of lineshapes, converging to a Lorentzian lineshape for $q=0$.
Fitting the resonance lineshape of the simulated reflectance spectrum with Eq.~\ref{eqn:fano} reveals that $E_{\text{c}}$ matches well the position of maximum Purcell enhancement, as can be seen in Figure~\ref{fig:1}(e). 

We now turn to the experimental results, which are obtained by monitoring the properties of different CBRs and embedded QDs upon repeated etch cycles of the native oxide.
The excitonic QD emission is centered at 1.581~eV~(784~nm) (standard deviation 8~meV~(4~nm)), whereas the CM positions before etching are 1.548(4)~meV~(801(2)~nm) for d1-CBRs, 1.542~meV~(804(2)~nm) for d2-CBRs, and 1.538~meV~(806(2)~nm) for d3-CBRs. 
One etch cycle consists of the growth of native oxide~\cite{Lukes1972,Reinhardt1996} by exposure to the ambient and the removal of that oxide by soaking the sample for 1 min in 18.5~\%~HCl~\cite{DeSalvo1996}.

\begin{figure}
  \includegraphics{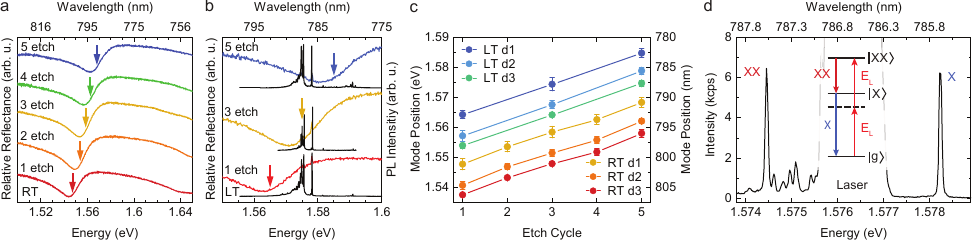}
  \caption{Relative reflectance spectra of a representative CBR, showing the etch-induced blue-shift of the CM at (a) room temperature (RT) and (b) low temperature (LT) (colored curves). Resonance positions $E_{\text{c}}$ are marked with an arrow. (b) also shows the PL spectrum (black curves) of the embedded QD obtained under above-bandgap excitation. (c) Mean values of $E_c$ for 10 CBRs having the same nominal design (either d1, d2, or d3)  at RT and LT as a function of the number of performed etch cycles. Error bars correspond to the standard deviation. (d) TPE spectrum with labeled X and XX transitions and remaining excitation laser in the background. Inset: Level scheme of $\ket{\text{XX}}$ population and radiative cascade decay.}
  \label{fig:2}
\end{figure}

Figure~\ref{fig:2}(a) shows the ratio of the reflected signal of an incident focused beam of thermal light on a representative CBR and on the surrounding planar regions, i.e., the relative reflectance, after each of the 5 performed etch cycles at room temperature (RT). 
All reflectance spectra are fitted with a Fano lineshape and the resonance positions $E_{\text{c}}$ are marked with an arrow. 
In order to quantify the mismatch of the emission to the CM, we record both the photoluminescence (PL) of the contained QD under above-bandgap excitation and the relative reflectance spectra at low temperature (LT), $\sim$10~K [see Figure~\ref{fig:2}(b)].
Due to cooling the resonators to LT, a blue-shift of the resonance position of 16.4(3)~meV (8.4(2)~nm) is observed.
The spectral position of the QD emission is not affected by the etching.
Figure~\ref{fig:2}(c) shows the mean values of the resonance positions of ten structures with the same nominal design (either d1, d2, or d3) for each etch cycle, revealing an average CM blue-shift of 5.1(2)~meV (2.6(1)~nm) per etch cycle.
The exposure time to ambient influences the amount of native oxide forming on the semiconductor surface, and consequently the magnitude of the CM shift.
The shift induced by the first etch cycle is more than twice as large than the average, as the time between the sample fabrication and the first etching was on the order of several days, whereas the exposure time of etch cycles 1, 2, 3, and 4 is $\sim$3 hours.
The CM shift induced by the fifth etch cycle is also larger than the average, i.e., 6.1~meV (3.1(2)~nm), since the exposure to ambient lasted one full day.
We expect the opposite behavior, and therefore, fine-tuning of $\sim$1~nm per etch cycle when exposing surfaces for tens of minutes to air~\cite{Sunner2007}.
The overall explored tuning range shows a total CM blue-shift of 31(3)~meV (16(2)~nm).
From a comparison between the measured and calculated shift based on the simulations we estimate that each etch step results in the removal of 0.9(3)~nm of material from each surface.

To analyze changes in the emission characteristics of on-demand emitters in tuned CBR structures, we employ a two-photon excitation (TPE) scheme, allowing us to access the decay times (lifetimes) of the biexciton $\ket{\text{XX}}$ and exciton $\ket{\text{X}}$ level in parallel.
The biexciton level $\ket{\text{XX}}$ is resonantly excited with a pulsed laser with a repetition rate of 80~MHz, where the laser energy $E_\text{L}$ is set to half of the energy difference between $\ket{\text{XX}}$ and ground state $\ket{\text{g}}$ ($E_{\text{XX}}$) and with the laser power adjusted to maximize the $\ket{\text{XX}}$ population~\cite{Muller2014}.
An additional above-bandgap light source is used to maximize the population efficiency and reduce QD blinking~\cite{Schimpf2019}.
A spectrum under such excitation conditions is provided in Figure~\ref{fig:2}(d), showing the transitions from $\ket{\text{XX}}$ to $\ket{\text{X}}$ (XX photons) and from $\ket{\text{X}}$ to $\ket{\text{g}}$ (X photons) in a cascade process.

\begin{figure}
  \includegraphics{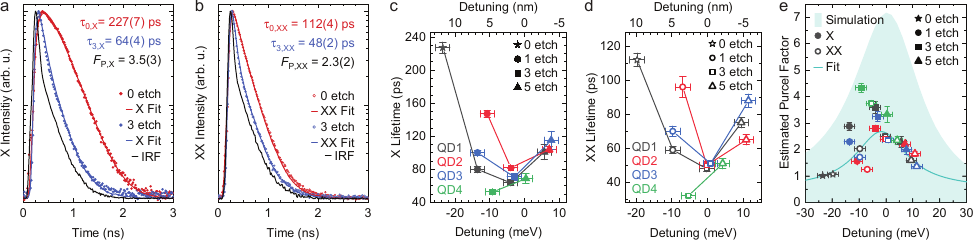}
  \caption{Time-correlated single-photon-counting measurements of X (a) and XX (b) photons emitted by QD1 upon TPE before (0 etch) and after 3 etch cycles. IRF denotes the instrument response function and the quoted lifetimes clearly indicate a lifetime reduction, which we attribute to the Purcell effect. (c) Analysis of lifetime values as a function of detuning from the cavity mode and different etch cycles for X and (d) XX. (e) Estimated Purcell factor as function of detuning, fitted with a Lorentzian function and compared to the simulated Purcell factor-spectrum. (c-e) Different colors are used for different QDs, as labeled in (c) and (d) while full/empty symbols are used for X/XX photons. Error bars (vertical) of the estimated Purcell factor are only based on the error of lifetime measurements. Error bars (horizontal) in the detuning axis are 2~meV, estimated experimentally from reflectance measurements. The detuning values of the star-data points were not directly measured but are estimations from the CM shift between etch cycle 0 and 1, measured using another CBR.}
  \label{fig:3}
\end{figure}

From Figure~\ref{fig:2}(b) we expect the highest coupling between an exciton confined in QD1 and the corresponding CM after etch cycle 3.
To confirm this, we performed time-correlated single-photon-counting measurements of X [XX] photons emitted by QD1, as displayed in Figure~\ref{fig:3}(a) [Figure~\ref{fig:3}(b)].
By comparing the decay dynamics of the untreated sample and after the 3\textsuperscript{rd} etch cycle, we observe a reduction of the lifetime $\tau$ for the $\ket{\text{X}}$ [$\ket{\text{XX}}$] states and estimate a Purcell enhancement $F_\text{P}$ of 3.5(3) [2.3(2)] based on $F_\text{P}=\tau_0/\tau_3$ where the index denotes the number of performed etch cycles.
Lifetime values $\tau_0$ of $\ket{\text{X}}$ and $\ket{\text{XX}}$ of QDs inside as-fabricated cavities coincide with values from similar QDs in the bulk (i.e., 230 ps for X and 120 ps for XX), as expected from the simulation, yielding $F_\text{P}\sim 1$ for such detuning.
We extend the study on decay dynamics to 4 QDs and present values for $\tau$ of X [XX] photons in Figure~\ref{fig:3}(c) [Figure~\ref{fig:3}(d)] as a function of detuning of the transition with the CM position, i.e., ($E_{\text{c}} - E_{\text{X/XX}}$). 
A clear reduction of the lifetime of all studied QDs is visible when approaching low detuning, followed by an increase when the CM is further blue-detuned. 
In order to compare the expected Purcell enhancement with the experimental data, we estimate the $F_\text{P}$ for the X and XX photons by using the typically measured lifetime values in bulk, which we quoted above. 
The results for all measured QD/CBR systems are plotted in Figure~\ref{fig:3}(e) as a function of detuning. 
The estimated $F_\text{P}$ data are accompanied by a Lorentzian fit, which is centered at -2(2)~meV, compatible with the expected 0~meV. 
The highest Purcell factor observed is 4.3(2).  
In spite of the fact that the measured and simulated cavity quality factor match well (see [Supporting Information]), the simulation predicts Purcell factors that are consistently higher than those extracted from the experiment (see shaded curve in Figure~\ref{fig:3}(e)). 
We attribute this observation and the pronounced scatter of the data points to the already mentioned spatial mismatch between emitters and cavity, since a radial misplacement exceeding $\sim$35~nm leads to a sub-optimal coupling even in case of spectral matching~\cite{Rickert2019}.

\begin{figure}
  \includegraphics{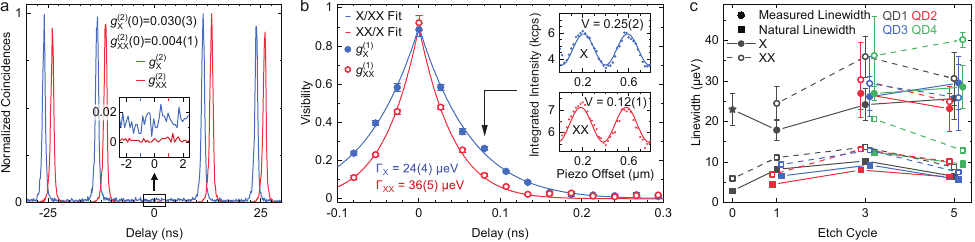}
  \caption{(a) $g^{(2)}(t)$ auto-correlation and (b) $g^{(1)}(t)$ coherence measurements of X (blue line) and XX (red line) photons for QD1 after (a) 3 and (b) 5 etch cycles. (a) XX/X peaks are horizontally shifted by +/-1~ns for ease of reading. (b) Insets show X and XX interference fringes at 80~ps time delay. (c) X (full/solid) and XX (empty/dashed) transition linewidth values of the studied QDs for different etch cycles, comparing measured linewidths (circle) to the expected natural linewidths (square) based on the respective lifetime. The star-data point was obtained under non-resonant excitation.}
  \label{fig:4}
\end{figure}

To probe the effect of etching on the multi-photon emission probability, we perform a Hanbury~Brown-Twiss (HBT) experiment after each etch cycle to measure the auto-correlation function $g^{(2)}(t)$ for time delays $t$ and evaluate it at $t=0$. 
The experimental results for the last etch cycle of QD1 are provided in Figure~\ref{fig:4}(a), yielding $g^{(2)}_{\text{X}}(0)=0.030(3)$ and $g^{(2)}_{\text{XX}}(0)=0.004(1)$. 
A strong suppression of the peaks at 0-time delay proves the single photon generation. 
As depicted in the zoomed-in panel, the $g^{(2)}_{\text{X}}(0)$ value of the X photons is somewhat higher than for XX photons, which we attribute to unintentional population of the $\ket{\text{X}}$ state by the  continuous-wave non-resonant laser used to reduce blinking. 
The $g^{(2)}(0)$ values for X and XX photons of QD4, which has the shortest lifetimes of the measured structures, are $g^{(2)}_{\text{X}}(0)=0.029(2)$ and $g^{(2)}_{\text{XX}}(0)=0.009(6)$, showing that the accelerated decay does not lead to re-excitation~\cite{Hanschke2018,Rota2022}.
Measurements of the auto-correlation function for both X and XX photons on 4 different QDs after etch cycles 1, 3, and 5 show that the values at 0-time delay are not affected by the etching (see section ``Auto-Correlation Data" in the [Supporting Information]). 

To gain further insight into the possible degradation of the optical quality of X and XX photons, we perform Michelson interferometry measurements to obtain the first-order coherence function $g^{(1)}(t)$, by probing the visibility of interference fringes for different time delays $t$ between the optical paths of the interferometer, as depicted in Figure~\ref{fig:4}(b). 
The potential influence of etching to linewidth broadening is determined by repeating $g^{(1)}(t)$ measurements on more QDs after different etch cycles. 
In Figure~\ref{fig:4}(c) the obtained linewidth values are shown and compared to the calculated natural linewidth $\hbar/\tau$ associated with the corresponding lifetimes of $\ket{\text{X}}$ and $\ket{\text{XX}}$, depicting the lower limit of sole lifetime broadening. 
The measurements indicate that no pronounced broadening of the linewidth can be observed within the error bars, even when applying 5 etch cycles, rendering this post-fabrication tuning method effective for broad-band tuning without deteriorating the optical quality of single photons. 

In summary, we have demonstrated that repeated wet-chemical etching and air exposure provides a simple and effective method to blue-shift the cavity modes of circular Bragg grating resonators in a spectral range of 31(3)~meV~(16(2)~nm).
Furthermore, this post-fabrication tuning allows to increase the emitter-cavity coupling in initially detuned systems, leaving the low multi-photon emission probability, as well as the high optical quality practically unaffected.
The highest value of Purcell enhancement that could be obtained within this study is $F_{\text{P}}=4.3(2)$, resulting in an excitonic lifetime of 53(2)~ps and a linewidth of 2.2(2) times the Fourier transform limit.
In our experiment, we expect the value of $F_{\text{P}}$ to be mostly limited by a misplacement of the QDs from the center of the cavity, resulting from an error during fabrication and causing partially polarized emission.
In principle, the investigated technique can be used to tune resonators in any solid-state emitter system that grows a native oxide, both broad-range and with a fine resolution, depending on the ambient-exposure time between etch cycles.
We expect this work to be beneficial to the community, serving as a tool to increase the yield of working samples and, therefore, accelerate the research on solid-state quantum emitters as resources in novel quantum technologies.

 \begin{suppinfo}

Supporting Information includes complementary data (quality factor, polarization of cavity modes, gas condensation, and $g^{(2)}(0)$ values) and methods (simulation, sample fabrication, measurement setup, and details of optical characterization) (PDF)

 \end{suppinfo}

\begin{acknowledgement}
The authors thank S. Bräuer, A. Halilovic and A. Schwarz from University of Linz and S. Kuhn from University of Würzburg for technical assistance.
This work was financially supported by the Austrian Science Fund (FWF) via the Research Group FG5, I~4320, I~4380, the European~Union’s Horizon~2020 research and innovation program under Grant Agreements No.~899814 (Qurope) and No.~871130 (Ascent+), the QuantERA~II~Programme that has received funding from the European~Union’s Horizon~2020 research and innovation programme under Grant Agreement No.~101017733 via the project QD-E-QKD and the FFG Grant No.~891366, the Linz~Institute~of~Technology (LIT), the LIT Secure and Correct Systems Lab, supported by the State of Upper Austria, the PNRR MUR project PE0000023-NQSTI (the National Quantum Science and Technology Institute), and the State of Bavaria.
\end{acknowledgement}

\bibliography{references}

\end{document}


\section{Complementary Data}
\subsection{Quality Factor}
\begin{figure}[htbp]
    \centering
    \includegraphics{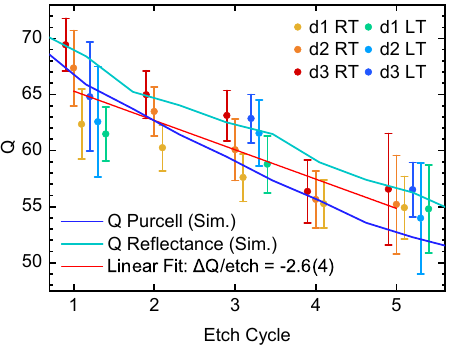}
     \caption[sfig:Qfactor]{Measured $Q$ factor as function of etch cycles, compared to values extracted from the simulation. Error bars are the standard deviation including a systematic error of $\pm2$, found from the uncertainty of $Q$ resulting from fitting reflectance spectra with different fit ranges. Data points are shifted horizontally around the respective etch cycle for better clarity.}
    \label{sfig:Qfactor}
\end{figure}

From the analysis of cavity mode (CM) positions upon repeated etching of circular Bragg grating resonators (CBRs), as discussed in the main text, the quality factor $Q$ can be extracted as well. 
$Q$ is defined as $Q=E_{\text{c}}/\Gamma_{\text{c}}$ with $E_{\text{c}}$ being the CM position, and $\Gamma_{\text{c}}$ the full width at half maximum (FWHM) of the resonance.
Supplementary~Figure~\ref{sfig:Qfactor} shows the mean values of $Q$, differentiating between CBR designs d1, d2, and d3, low temperature (LT), and room temperature (RT), corresponding to the same CBRs analyzed in Figure~2(c) in the main text.
In the experiment we see a slight decrease of $Q$ values for increasing etch cycles, which is fully consistent with the calculation results. 
To compare measured $Q$ values with the simulation, the abscissae require to be matched with each other.
As it was already presented in the main text, the material removal depth used in the simulation is 1.5~nm per etching, whereas the experimentally obtained value is 0.9~nm per etch cycle.
A linear fit of the measured values yields a decrease of $Q$ of 2.6(4) per etch cycle. 

\subsection{Polarization-Resolved Measurements}
Polarization-resolved measurements of light reflected from the cavity, as well as photoluminescence (PL) signal from measured quantum dots (QDs) is performed by turning a half-wave plate in front of a polarizer step-wise by the angle $\Delta \theta$ in a range of 180\textdegree, recording a spectrum after each turning step.
Polarized features in such a spectrum series are oscillating as function of 2$\theta$.
As mentioned in the main text, QDs exhibit polarized emission, where the majority shows a polarization degree between 60~\% and 99~\% and a common polarization axis within 8\textdegree standard deviation.

We also measure a polarization-splitting of horizontal (H) and vertical (V) resonator modes with a magnitude of about 1~nm~(1.9~meV) among the studied CBRs.
Relative reflectance spectra of such modes of a representative CBR are given in Supplementary~Figure~\ref{sfig:refl_pol}.
Due to the large overlap of H and V modes, we attribute no substantial effect of this splitting on the polarized emission of the QD. 

\begin{figure}[htbp]
    \centering
    \includegraphics{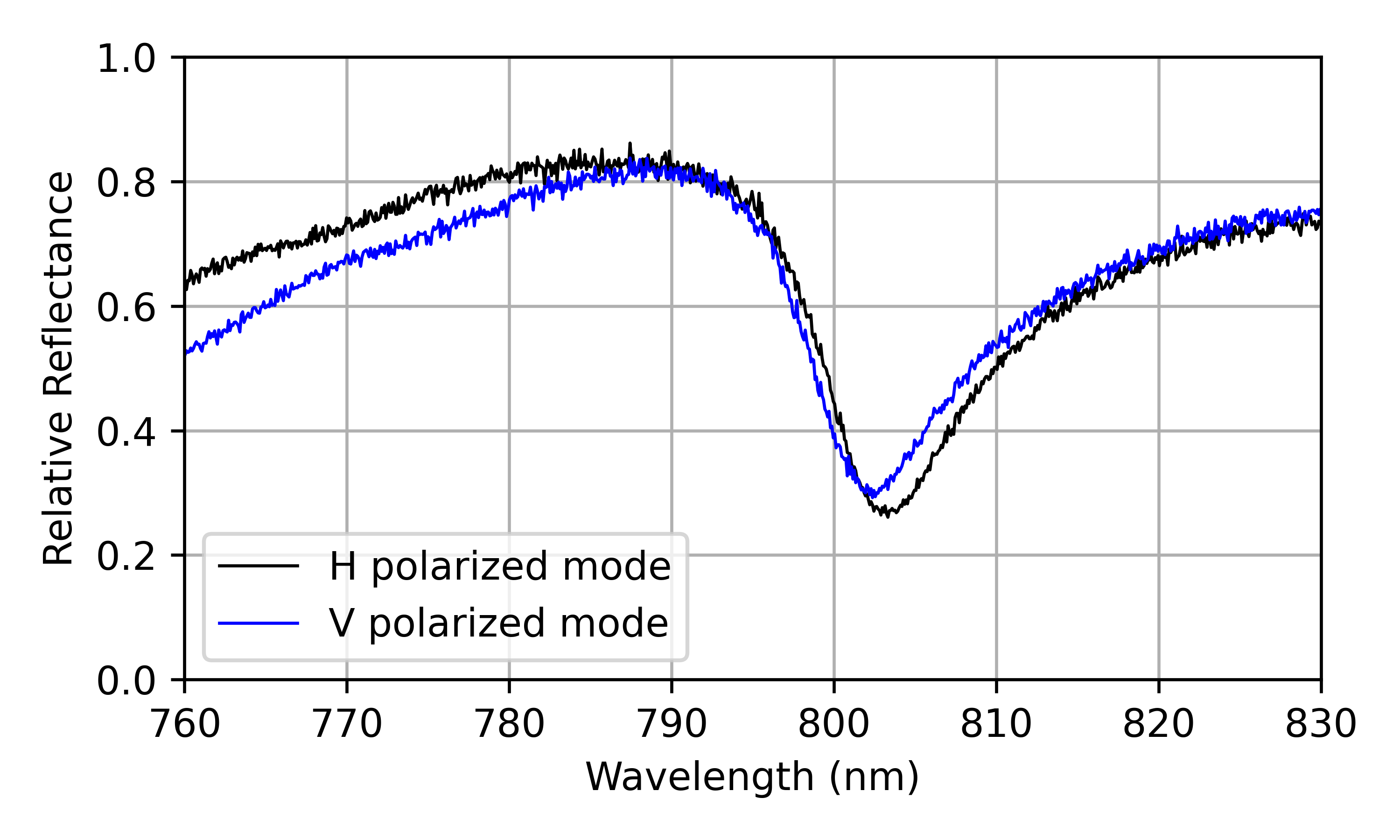}
    \caption[sfig:reflpol]{Relative reflectance of H and V polarized modes shifted 1~nm~(1.9~meV) with respect to each other. The resonance position of the H~polarized mode is at 801.7~nm~(1.5465~eV), whereas the V~polarized mode is located at 800.7~nm~(1.5484~eV).}
    \label{sfig:refl_pol}
\end{figure}

\subsection{Gas Condensation}
When measuring CM positions of CBRs at low temperature ($<8$~K) for etch cycle 3, the CM did not shift as anticipated.
While we were expecting a temperature-induced blue-shift of 8.4(2)~nm (as seen in the main text), we observed a smaller blue-shift of 6(2)~nm instead.
The pressure during this measurement was $\sim 10^{-4}$~mbar, compared to a value $<10^{-6}$~mbar that we usually achieve.
After careful inspection of the vacuum system, the experiment was repeated on another day with a pressure of $<10^{-6}$~mbar, resulting in a temperature-induced blue-shift of the CM of 8.2(2)~nm.
The difference of the CM positions between the first and second measurement (different pressure in the cryostat) at low temperature is 3(2)~nm, whereas the CMs do not show any shift at room temperature. 
We attribute the observed behaviour to gas condensation on the CBRs~\cite{Mosor2005}. 
Another indicator of gas condensation on the sample surface is the observation of an increasing amount of visible contamination at low temperature at a pressure of $\sim 10^{-3}$~mbar. 
Dark spots appear and cover the sample surface and can be removed by heating the sample to room temperature.
An image of such contamination is given in Supplementary~Figure~\ref{sfig:refl_gas_cond}.
(Note that the quoted pressure values were measured in proximity to a turbomolecular pump, connected to the cryostat with an approximately 2.5~m long tube)

\begin{figure}[htbp]
    \centering
    \includegraphics[scale=0.5]{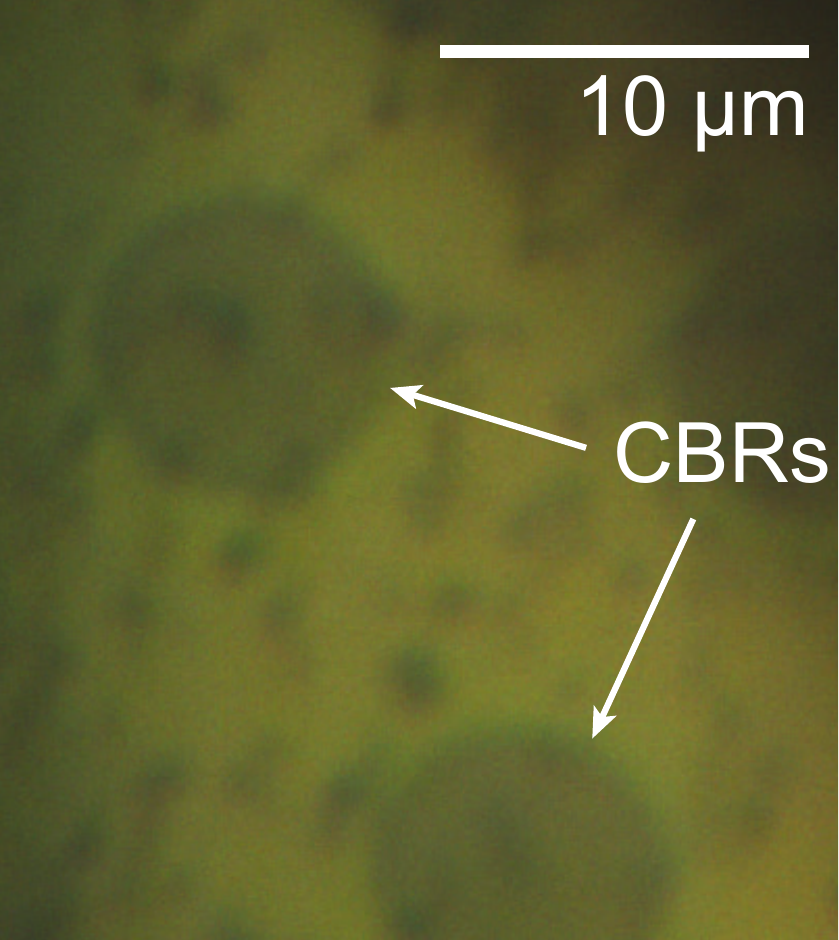}
    \caption{A wide-field image of the sample surface at a pressure of $\sim 10^{-3}$~mbar, showing CBRs and dark spots distributed over the sample.}
    \label{sfig:refl_gas_cond}
\end{figure}

\subsection{Auto-Correlation Data}

The measured data set of values of $g^2(t)$ at 0-time delay for different etch cycles of the studied QDs is provided for X and XX photons in Table~\ref{tab:g2}. 
Due to fast blinking dynamics of the source, only the first neighboring peaks are taken into account for normalization. The used bin-width is 2~ns. 
Increased values of $g^{(2)}(0)$ of X photons are due to a background originating from non-resonant excitation, as discussed in the main text, and do not correlate with the repetition of etch cycles.
Details about the measurement are given below in the subsection~``Auto-Correlation". 

\begin{table}[htbp]
\caption[tab:g2]{Values of $g^{(2)}(0)$ for different etch steps. The highlighted row in green corresponds to the measurement given in the main text in the inset of Figure~4(a) and the one in teal corresponds to the QD with the shortest lifetimes of $\ket{\text{X}}$ and $\ket{\text{XX}}$.}
\begin{tabular}{|l|l|l|l|}
\hline
QD No. & Etch Cycle & $g^{(2)}(0)_{\text{X}}$ & $g^{(2)}(0)_{\text{XX}}$ \\ \hline
1      & 0          & 0.059(6)                & -                         \\ \hline
1      & 1          & 0.024(2)                & 0.025(3)                 \\ \hline
1      & 3          & 0.072(2)                & 0.007(1)                 \\ \hline
\cellcolor[HTML]{95fc9c}1      & \cellcolor[HTML]{95fc9c}5          & \cellcolor[HTML]{95fc9c}0.030(3)                & \cellcolor[HTML]{95fc9c}0.004(1)                 \\ \hline
2      & 3          & 0.051(4)                & 0.012(2)                 \\ \hline
2      & 5          & 0.093(9)                & 0.006(2)                 \\ \hline
3      & 3          & 0.017(2)                & 0.005(1)                 \\ \hline
3      & 5          & 0.096(5)                & 0.049(4)                 \\ \hline
\cellcolor[HTML]{32a0a8}4      & \cellcolor[HTML]{32a0a8}3          & \cellcolor[HTML]{32a0a8}0.029(2)                & \cellcolor[HTML]{32a0a8}0.009(6)                 \\ \hline
4      & 5          & 0.073(4)                & 0.014(2)                 \\ \hline
\end{tabular}
\label{tab:g2}
\end{table}

\section{Methods}

\subsection{Simulation}

The 3D finite-difference time-domain (FDTD) simulations are performed using the commercial solution ANSYS Lumerical. 
The CBR is centered in the simulation domain with the parameters periodicity $p=380$~nm, trench width $t=100$~nm, central disc radius $r=333$~nm, membrane thickness $d=148$~nm, and oxide thickness $d=200$~nm, stacked on a gold substrate.
The used refractive index of the membrane is $n_{\text{AlGaAs}}=3.3$ and of the oxide $n_{\text{Al$_{2}$O$_{3}$}}=1.64$. The implemented refractive index of gold is wavelength-dependent~\cite{johnson1972optical}.
For each simulated etch cycle, 1.5~nm of material is removed, i.e., $t$ [$d$ and $r$] increases [decreases] 1.5~nm per step.

The relative reflectance is defined as the division of the reflectance of the CBR by the reflectance of the layer stack without CBR, to resemble the experiment. 
Therefore, in the simulations, a Gaussian beam with beam radius of 750~nm and a divergence angle of 30° is focused on the center of the CBR. The beam central wavelength is 800~nm with a range of 200~nm.
The reflected intensities are recorded with a numerical aperture of 0.65. 
As the polarization of the Gaussian source is oriented along the $x$-axis, simulation time is reduced by using antisymmetric perfectly matched layer (PML) boundary conditions (BCs) normal to the $x$-axis and symmetric PML BCs normal to the $y$-axis. 
Due to the bottom gold mirror, metal BCs are used at the bottom of the simulation domain, whereas PML BCs are used at the top.

For the simulations targeting extraction efficiency and Purcell factor, a dipole source with a central emission wavelength of 780~nm and a range of 160~nm is used. 
The dipole emitter is oriented along the $x$-axis, allowing the above-mentioned symmetry and BCs to be used again. 
The extraction efficiency in the far-field can be calculated from a monitor placed above the structure in extraction direction\cite{https://doi.org/10.1002/adom.202001490}. 
For this
\begin{equation}
    \eta_{ex}(\theta)=\frac{\int_{0}^{2\pi}\int_{0}^{\theta}\left|E_\lambda(\theta',\phi)\right|^2\sin{(\theta')}\text{d}\theta' \text{d}\phi}
    {\int_{0}^{2\pi}\int_{0}^{\pi/2}\left|E_\lambda(\theta',\phi)\right|^2\sin{(\theta')}\text{d}\theta' \text{d}\phi}\cdot\frac{T}{F_{\text{P}}},
\end{equation}
can be used, where $E_\lambda$ is the electric far-field depending on the wavelength $\lambda$, $F_{\text{P}}$ the Purcell factor and $T$ the near-field transmittance, integrated over spherical coordinates $\theta,$ and $\phi$. 

The Purcell factor is calculated by the relation\cite{novotny_hecht_2012}

\begin{equation}
    \frac{\Gamma}{\Gamma_0}=\frac{P}{P_0}=F_{\text{P}},
\end{equation}

where $P$ is the enhanced energy dissipation derived from Poynting's theorem, $\Gamma$ is the enhanced transition rate and $P_0$ and $\Gamma_0$ are the respective reference values of a source in bulk. 
Consequently, the Purcell factor is obtained from the simulations by placing monitors around the source which detect the transmission of radiated source power.

\subsection{Sample Fabrication}
QD samples used within this study were grown by  molecular beam epitaxy (MBE).
To assist the subsequent resonator fabrication, first, a 400~nm thick etch stop layer of Al$_{0.75}$Ga$_{0.25}$As is grown on top of a commercial GaAs(001) wafer following GaAs-buffer growth. Then, a 4~nm GaAs  layer and a 70~nm Al$_{0.33}$Ga$_{0.67}$As barrier layer is deposited. 
Utilizing the local etching of Al droplets, symmetric nanoholes are formed and filled with 2.5~nm GaAs, overgrown with 69~nm Al$_{0.33}$Ga$_{0.67}$As and capped with 4~nm of GaAs, producing a sample with strain-free GaAs QDs.

\begin{figure}
  \includegraphics{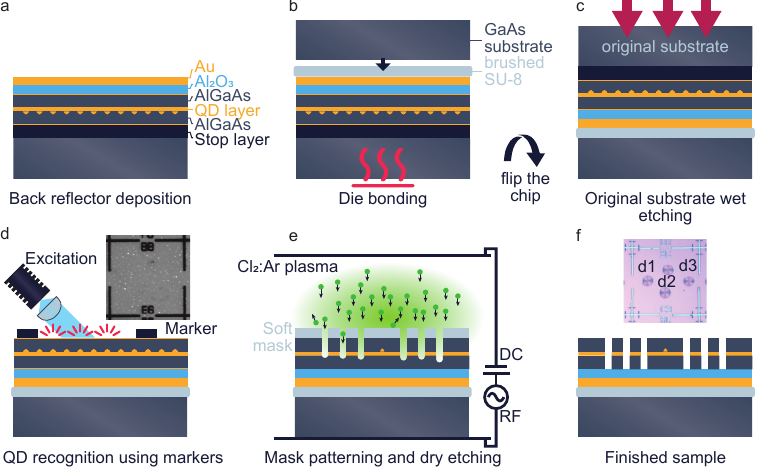}
  \caption{Sample fabrication, consisting of (a) the deposition of Al$_2$O$_3$ and Au, (b) bonding to a new substrate using SU-8, (c) multi-step wet-chemical etching of the original substrate, (d) cryogenic wide-field imaging and numeric image processing to map QD position, and (e) deterministic patterning and etching of CBRs, providing (f) a finished sample of QD-CBR devices.}
  \label{sfig:fabrication}
\end{figure}

A chart of the processing flow to obtain reflector-backed membranes is depicted in Supplementary~Figure~\ref{sfig:fabrication}(a-c).
Pieces of approximately 10~mm$^2$ size are cut from the wafer and covered with 200~nm of Al$_2$O$_3$ by atomic layer deposition (ALD), a 2~nm layer of e-beam evaporated Cr and 150~nm of thermally evaporated Au, forming the back reflector of the cavity.
Then, SU-8~2 is spread using a brush on the sample surface, dried and heated to 130~\textdegree C when the sample is brought to contact with a same-size GaAs substrate. 
Crosslinking of polymers is initiated by applying a pressure of 2.2~MPa at a temperature of 230~\textdegree C for 10~min.
Wet-chemical back-etching of the original substrate involves a fast etch system relying on the 3:7 mixture of 85~\%~H$_3$PO$_4$ and 30~\%~H$_2$O$_2$, stopping well before reaching the etch stop layer, and a 1:4 solution of 30~\%~H$_2$O$_2$ and powdered C$_6$H$_8$O$_7$ dissolved 1:1 in H$_2$O.
The latter solution is selective between the substrate and etch stop material, leaving this layer to be removed with 10~\% HF. 

In order to introduce a frame of reference for recording QD positions, metallic reference markers are deposited on the surface of the back-etched sample.
Electron-beam lithography (EBL) with 30~kV acceleration voltage is used to pattern marker crosses into the CSAR~62 e-beam resist, with an additional protective coating of Electra~92 to avoid charging effects.
After the resist development, the sample is covered with a 150~nm stack of equally thick and strain-compensated Cr-Au-Cr layers.
The lift-off in acetone and anisole reveals the marker fields defined by the remaining crosses and labels. 
To find QD positions, the sample is placed into a liquid He continuous-flow cryostat, the surface is illuminated with an infrared light-emitting diode (LED), whereas QDs are excited with a blue LED, making them visible through PL~\cite{Krieger2023}. 
Images of the sample with centered marker fields are processed numerically, to find QD positions, i.e., the center of fitted 2D-Gaussian shapes, with respect to the coordinate system, found by fitting the marker crosses.
A sketch of the proceeding and an example of image can be seen in Supplementary~Figure~\ref{sfig:fabrication}(d). 

\begin{figure}
  \includegraphics{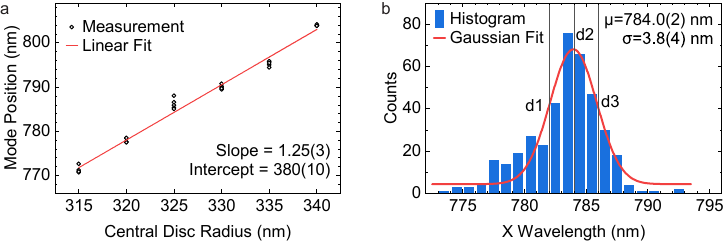}
  \caption{(a) Calibration curve to control the resonance of the CBR with the central disc radius $r$. (b) Histogram of the X wavelength fitted with a Gaussian with mean $\mu$ and standard deviation $\sigma$. Chosen CBR designs d1, d2, and d3 are marked.}
  \label{sfig:calibration}
\end{figure}

To match the CBR spectrally with the QD emission, a set of empty resonators is etched into the sample and mode positions as a function of central disc radius $r$ are characterized, as depicted in Supplementary~Figure~\ref{sfig:calibration}(a).
Since the spectral position of the exciton photon X is not recorded for single QDs but for the ensemble only, three designs d1, d2, and d3 are chosen to increase chances of spectrally matching the emission with the CM. A histogram of the emission statistics of this sample with the chosen target resonances at 782~nm~(1.585~eV), 784~nm~(1.581~eV), and 786~nm~(1.577~eV) is provided in Supplementary~Figure~\ref{sfig:calibration}(b).
With known QD positions and a calibrated $r$ parameter, CBRs can be fabricated deterministically, using 100~kV EBL and reactive-ion etching with Cl$_2$:Ar chemistry in an inductively-coupled plasma reactor, yielding the finished sample, see Supplementary~Figure~\ref{sfig:fabrication}(e-f).

As mentioned in the main text, measurements indicate limited spectral and spatial overlap of the CBR with the embedded emitters. 
Due to a mistake in the design for deterministic EBL patterning, the actual realized parameters of the CBRs are periodicity $p=380$~nm, trench width $t=100$~nm, and radii $r_1=331$~nm,  $r_2=332.6$~nm, $r_3=334.2$~nm, for designs d1, d2, and d3, resulting in cavities detuned red with respect to the QD emission.
Furthermore, due to a rounding error in the script for detecting QD positions, QDs are off-center of the CBR, leading to polarized emission\cite{Peniakov2023}. 

\subsection{Optical Characterization}
The sample is repeatedly measured between different etch cycles using optical characterization techniques in a confocal setup, sketched in Supplementary~Figure~\ref{sfig:setup}. 
A green 532~nm diode laser is used in continuous-wave (CW) operation to excite the QDs in the above-bandgap regime, whereas a wavelength-tunable pulsed titanium-sapphire (TiSa) laser enables on-demand photon generation from QDs (see below).
Separate white light sources are used for widefield imaging and broadband reflectance measurements (see below).
The polarization state of the light can be prepared with a polarizer (POL) and a rotating half-waveplate (HWP) and is directed to the sample, residing in a liquid-He continuous-flow cryostat, as can be seen in Supplementary~Figure~\ref{sfig:setup}(b).

Polarization-resolved spectroscopy, imaging, as well as correlation and interferometry measurements can be performed plug-and-play by distributing the collected signal with single-mode (SM) fibers, flippable mirrors (flip-M), and flippable beam splitters (flip-BS) as depicted in Supplementary~Figure~\ref{sfig:setup}(a).
To filter stray light from the room, a long-pass filter (LPF) is placed in front of the entrance slit of the spectrometer.
The exit slit of the single [double] spectrometer enables it to be used as a monochromator, necessary for correlation experiments, depicted in Supplementary~Figure~\ref{sfig:setup}(c)[(d)].
Avalanche photodiodes (APDs) are used to detect single photons, connected to correlation electronics (see below).
In Supplementary~Figure~\ref{sfig:setup}(e), a sketch of the Michelson interferometer shows a linear stage with a mounted retroreflector (RR) moving between interference visibility measurements using another RR on a piezoelectric actuator (see below).

\begin{figure}
  \includegraphics{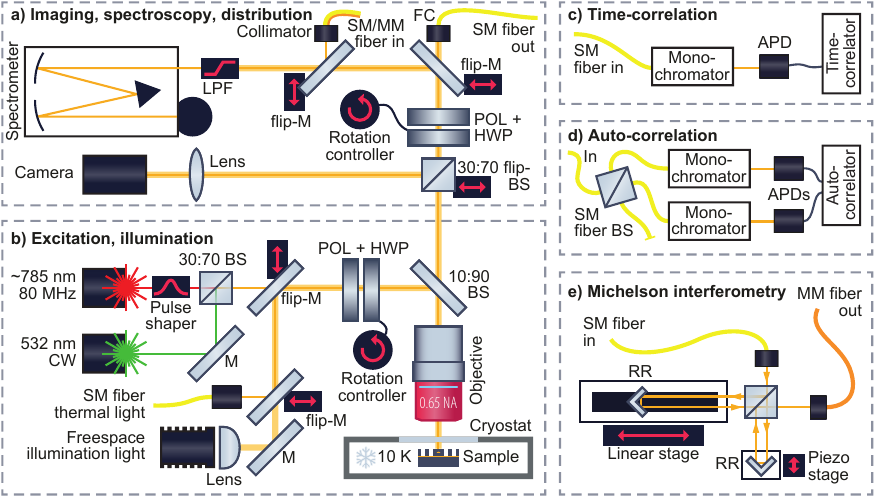}
  \caption{Sketch of the setup with highlighted dedicated areas.}
  \label{sfig:setup}
\end{figure}
\subsubsection{Reflectance Measurements}

Mode measurements of CBRs are conducted using reflectance measurements, where a thermal light from a halogen source is coupled into a SM fiber and directed to the sample. Reflected light is collected using a spectrometer with a 300 lines/mm diffraction grating, featuring a resolution of 0.21~nm, a vertical binning of 10 pixel (each 20~$\times$~20~\textmu m$^2$) of the charge-coupled device (CCD) to reduce noise contribution, and an entrance slit of 80~\textmu m or larger.
Obtained reflectance spectra are normalized by the reflected signal of the un-etched membrane surrounding the CBR structures.
From experimental experience we observe that the reflectance minimum position is slightly dependent on the polarization (see above), as well as on the exact spatial position of the beam on the surface. 
Therefore, we conclude an experimental error of $\pm$1~nm of the mode position.

\subsubsection{Two-Photon Excitation}

In order to deterministically pump the biexciton state $\ket{\text{XX}}$, two-photon-excitation (TPE) is used. To achieve this, a TiSa laser, producing pulses with a repetition rate of 80~MHz and a pulse length of about 100~fs, stretched to $\sim$5~ps using a pulse shaper, is employed.
To excite QDs showing highly-polarized emission due to spatial misplacement with the cavity, it is necessary to match the polarization state of the excitation laser with the QD emission.
The energy of the laser light is tuned to half of the energy $E_{\text{XX}}$, i.e., the distance between $\ket{\text{XX}}$ and the ground state $\ket{\text{g}}$.
$\ket{\text{XX}}$ shows a binding energy $E_{\text{b}}$ with respect to the exciton state $\ket{\text{X}}$:

\begin{equation}
    E_{\text{XX}}/2 = E_{\text{X}}-E_{\text{b}}/2
\end{equation}

For GaAs~QDs this binding energy has found to be always positive and fairly constant at $E_b\approx$3.8~meV(2~nm). 
Since $\ket{\text{XX}}$ is driven resonantly, the state population experiences phonon-damped Rabi oscillations~\cite{PhysRevLett.91.127401,PhysRevB.73.125304} as a function of laser power. All measurements are conducted on the global maximum of the Rabi oscillations, the so called $\pi$-pulse.

\subsubsection{Time-Correlation}

Measurements of the lifetime of $\ket{\text{X}}$ and $\ket{\text{XX}}$ states are performed by time-correlated single photon counting.
As mentioned before, emission into a polarized mode is favored on this sample; therefore, the polarization axis of the TPE pulse is aligned with respect to that axis.
Furthermore, the emission is filtered with a rotated half-waveplate and a polarizer, such that only photons with a polarization state aligned to that axis are measured, as we expect the highest emitter--cavity coupling for this case.
Emitted photons are collected using a SM fiber, filtered for X/XX signal using the exit slit of a spectrometer and forwarded to an APD with a time resolution of $\sim100$~ps, as depicted in Supplementary~Figure~\ref{sfig:setup}(c).
To obtain more accurate results of lifetime values, the instrument response function (IRF) is recorded, by sending attenuated laser signal to the detector.
For extracting the lifetimes of $\ket{\text{X}}$ and $\ket{\text{XX}}$ states, the histograms of the arrival time of the corresponding photons on an APD, relative to a clock from the pulsed excitation laser are fitted. 
Here we give a short explanation of the fitting functions.

From the rate equations, one can follow, that after the population of $\ket{\text{XX}}$, there should follow an exponential decay from $\ket{\text{XX}}$ to $\ket{\text{X}}$ followed by a bi-exponential decay from $\ket{\text{X}}$ to $\ket{\text{g}}$. 
Therefore, the population as a function of time for the two states, corresponding to the time-dependent intensity profile $I_{\text{X/XX}}(t)$ should have the following form:

\begin{align}
    I&(t)_{\text{XX}}(t)=A\exp(-\frac{t}{\tau_{\text{XX}}})\\
    I&(t)_{\text{X}}(t)=B\exp (-\frac{t}{\tau_{\text{X}}}) + C\exp (-\frac{t}{\tau_{\text{XX}}}),
\end{align}
with $\tau_{\text{XX}}$ and $\tau_{\text{X}}$ being the lifetime of the XX/X respectively.
The fitting function used to extract the lifetimes is the convolution of the exponential/bi-exponential decay with the measured IRF.

\subsubsection{Auto-Correlation}
The measurement of the second-order auto-correlation function $g^{(2)}(t)$ is realized in a Hanburry Brown-Twiss experiment.
QD emission is coupled into a SM fiber, sent through a 50:50 fiber BS with the two outputs connected to two spectrometers used as monochromators, and forwarded to APDs, connected with correlation electronics, as sketched in Supplementary~Figure~\ref{sfig:setup}(d).
In the obtained correlation histogram, as seen in the main text, coincidences of detection events on both detectors are plotted as a function of the time delay between the detection of those two photons.
The antibunching nature of single photons reveals a dip at 0-time delay, indicating a reduced probability of detecting two photons simultaneously.
The coincidence peaks of photons originating from different excitation cycles are separated by 12.5~ns corresponding to the 80~MHz repetition rate of the excitation laser.

The second-order auto-correlation function is evaluated at 0-time delay, by integrating 2~ns around the 0-time delay dip divided by the mean of the two closest neighboring peaks, also integrated 2~ns.
Further side peaks are neglected, due to fast blinking dynamics of the QD (telegraph noise).
A method to reduce blinking is to assist the TPE with a green CW laser to maximize XX photon brightness; however, this degrades the $g^2(0)$ value of X photons, due to non-resonant undesirable excitation events of $\ket{\text{X}}$.

\subsubsection{Michelson Interferometry}

Michelson interferometry, i.e., the measurement of the first-order coherence $g^{(1)}(t)$, is used to determine the coherence properties of a light source, and thereby, further quantify its optical quality. 
The main components of the interferometer are a 50:50 BS, a retroreflector (RR) on a linear stage and a RR on a piezo stage. 
The incoming beam is split into two ideally identical parts, which are then reflected back onto the BS to interfere there. 
The piezo performs several steps of 20~nm for each position on the linear stage. 
There, the step size is chosen according to the expected coherence length in the range of a few mm.
A sketch of a Michelson interferometer is given in Supplementary~Figure~\ref{sfig:setup}(e).
The intensity of one of the outputs of the BS is measured using a spectrometer. 
For each position of the linear stage, the intensity of the emission line of interest is integrated which leads to sets of intensity data where the interference is visible for different relative delay times. 
These are fitted using a cosine function, to extract the visibility $\nu$ at each of a given time delay:
\begin{equation}\label{equ::vis}
    \nu=\frac{I_{max}-I_{min}}{I_{max}+I_{min}}
\end{equation}\\
This is valid, if we assume that the optical path difference created by the piezo stage is very small compared to the coherence length and therefore the visibility is constant in that range. 
This visibility in dependence of the relative delay time is fitted using the Fourier-transform (FT) of a Voigt function, which we explain now in more detail:

The lower limit on the radiative linewidth is given by the lifetime $\tau_{\text{X/XX}}$ of the exponential decay of $\ket{\text{X}}$ and $\ket{\text{XX}}$,leading to a Lorentzian line shape with a FWHM of $\Gamma_{0,\text{X/XX}} = \hbar/\tau_{\text{X/XX}}$, which is called the natural linewidth.
QD emission lines are broadened by spectral wandering, which occurs when there are charge fluctuations in the vicinity of the QD~\cite{Kamada2008,Kuhlmann2013}. 
This leads to a Gaussian broadening on macroscopic timescales, therefore in a Michelson measurement, the linewidth measured is the convolution of both effects:

\begin{align}\label{equ::voigt}
    V(E,\sigma,\gamma)&=(G\ast L)(E) = \int G(\epsilon)L(\epsilon-E)d\epsilon,\\
    G(E,\sigma)&=\frac{1}{\sigma\sqrt{2\pi}}e^{-\frac{E^2}{2\sigma^2}},\\
    L(E,\gamma)&=\frac \gamma {\pi\left(E^2+\gamma^2\right)},
\end{align}
with the Voigt profile $V(E)$, the Gaussian profile $G(E)$, and the Lorentzian profile $L(E)$ as function of photon energy $E$, and widths of the distributions $\sigma$ and $\gamma$.
The FT, which transforms the energy profile into the time profile where the Michelson measurement takes place, simplifies this to a multiplication, which is used for fitting the visibility data:

\begin{equation}\label{equ::voigt_ft}
    \text{FT}[V(E,\sigma,\gamma)] =  \hat{V}(t,t_G,t_L) =  \hat{G}(t,t_G)\cdot \hat{L}(t,t_L),
\end{equation}
with the coherence-times $t_G$ and $t_L$ for the Gaussian, and the Lorentzian part, respectively, and $\hat{V}$, $\hat{G}$, and $\hat{L}$, the FT of the corresponding profiles. 
The widths $\sigma,\gamma$ in the energy picture can be calculated using:

\begin{equation}\label{equ::sigma_gamma}
    \sigma = \frac{\hbar}{t_{G}}\quad \text{and} \quad \gamma = \frac{\hbar}{t_{L}}.
\end{equation}

In order to calculate the FWHM $f_V$ of the Voigt profile, the FWHM values $f_{G,L}$ are needed:
\begin{align}
    f_G&=2\sigma\sqrt{2\ln{2}}\\
    f_L&=2\gamma
\end{align}

Since the Voigt profile does not have an analytical form, the width can not be calculated analytically but there is an approximation~\cite{Olivero1977}:
\begin{equation}\label{equ::v_width}
    f_V \approx 0.5346f_L+\sqrt{0.2166f_L^2+f_G^2}
\end{equation}
For a purely Gaussian line, this approximation fits perfectly, while for an arbitrary Voigt profile it produces results with an accuracy of around $0.02\%$. 
Widths $f_V$ correspond to the measured linewidths provided in the main text.

\bibliography{supportingReferences}